\documentclass[twocolumn,showpacs,preprintnumbers,amsmath,amssymb]{revtex4}
\usepackage{graphicx}

\def\dd{\displaystyle}

\begin{document}

\title{\bf Radiative Correction to the Dirichlet Casimir Energy for $\lambda \phi^{4}$ Theory in Two Spatial Dimensions}
\author{S.S. Gousheh}%
\email{email: ss-gousheh@sbu.ac.ir}
\author{R. Moazzemi}%
\email{email: r-moazzemi@sbu.ac.ir}
\author{M.A. Valuyan}%
\email{email: m-valuyan@sbu.ac.ir}
\affiliation{%
Department of Physics, Shahid Beheshti University, Evin, Tehran
19839, Iran
}
\date{\today}
\begin{abstract}
In this paper, we calculate the next to the leading order Casimir
energy for real massive and massless scalar fields within
$\lambda\phi^{4}$ theory, confined between two parallel plates with
the Dirichlet boundary condition in two spatial dimensions. Our
results are finite in both cases, in sharp contrast to the infinite
result reported previously for the massless case. In this paper we
use a renormalization procedure introduced earlier, which naturally
incorporates the boundary conditions. As a result our radiative
correction term is different from the previously calculated value.
We further use a regularization procedure which help us to obtain
the finite results without resorting to any analytic continuation
techniques.

\end{abstract}
\maketitle
\section{Introduction}
The Casimir effect can be observed in all systems with nontrivial
boundary conditions\,(BCs) or background fields\,(\emph{e.g.}
solitons). During the last sixty years this effect has been an
important topic of research with applications in many branches of
physics\,\cite{plunien.,mostepanenko.,phys.rep.353}. The static
Casimir effect, first calculated in 1948\,\cite{h.b.g.}, predicts an
attraction between two perfectly conducting parallel plates, due to
distortion of the electromagnetic vacuum state\,(for a general
review on the Casimir effect, see
Refs.\,\cite{mostepanenko.,milton.paper.,milton.book.}). Ten years
later the first attempt to observe this phenomena was made by
Sparnaay \cite{sparnaay}. Since then, many experimental
investigations have measured precisely the Casimir force in
different cases, such as two parallel plates\,\cite{Bressi.}, or a
sphere in front of a plane\,\cite{Lamoreaux.decca1.decca2.}. The
majority of the theoretical investigations related to the zero order
Casimir effects are for various fields, geometries,
BCs\,\cite{milton.book.,wolf.,svaiters,cognola.}, and various
dimensions. Some of the major approaches used are: the mode
summation method with a combination of the zeta function
regularization technique\,\cite{mode.sum.,nesterenko.piro.}, Green's
function formalism\,\cite{green.func.}, multiple-scattering
expansions\,\cite{multiple.scatt.exp.}, heat-kernel
series\,\cite{heat.kernel.}. On the other hand, there exist many
works on the first order and also second order radiative corrections
to the Casimir energy for various
cases\,\cite{onetwo.loop.,6.reza.3,7va8.reza.3,cavalcanti.}. Some of
the major approaches used for the radiative corrections to the
Casimir effect are the phase shift of the scattering
states\,\cite{graham.IJMP.}, or the replacement of the BCs by an
appropriate potential term\,\cite{graham.jaffe.,milton.potential.}.
The Casimir effects have found many applications in physics. For
example, the Casimir effect is the major contribution to the
radiative correction to the mass of the solitons, and these
corrections have been investigated in many
papers\,\cite{vega...,dashen.,yamagishi.}.
\par
The value of the Casimir energy has a complicated behavior as a
function of the number of spatial dimensions, the type of fields,
type of topology, and geometry. The case of even spatial dimensions
is usually more
complicated\,\cite{cognola.,bender.milton.,cavalcanti.}. Since many
interesting condensed matter systems are well-approximated by two
dimensional models, extracting finite results from the complicated
divergencies, which usually plague such systems, is very important.
The Casimir energies for scalar fields in even dimensions have been
discussed for two parallel plates\,\cite{cavalcanti.},
spheres\,\cite{cognola.,bender.milton.}, and
cylinders\,\cite{nesterenko.piro.}. Some of those cases give
divergent results, and some authors prescribe methods to extract a
finite answer from those
expressions\,\cite{cognola.,milton.book.,nesterenko.piro.}. However,
those prescriptions are not universally applicable to all even
dimensions. It seems that even for the simplest case of a scalar
field in $2+1$ dimensions, it is not clear how the divergences can
be removed\,\cite{milton.book.}.
\par
In this paper we calculate the first order radiative correction to
the Dirichlet Casimir energy for two infinite parallel plates for
massive and massless scalar fields in two spatial dimensions. The
problems mentioned before, give us extra motivation to utilize an
alternative renormalization program and regularization procedure for
this problem. We have used these procedures to calculate this
quantity in $1+1$ and $3+1$ dimensions\,\cite{reza1.,reza2.}. As we
shall see our procedure yields finite results for both massive and
massless scalar fields which is different from the previously
reported one\,\cite{cavalcanti.}. As a matter of fact the previously
reported result for the massless case is
infinite\,\cite{cavalcanti.}. It is worth mentioned that our finite
result is obtained without any use of analytic continuation
techniques due to our regularization procedure. The difference
between our results and the previously reported one can be
attributed to our alternative renormalization program.
\par
In our paper we combine two independent programs in order to
calculate the radiative correction to the Casimir energy. First, we
use an approach to the renormalization program which we believe to
be systematic. The procedure to deduce the counterterms from the
$n$-point functions in the renormalized perturbation theory is
standard and has been available for over half a
century\,\cite{peskin.book.}. We believe that all of the information
about the nontrivial BCs or position dependent background fields
should be carried by full set of the $n$-point functions. Therefore,
all of the counterterms deduced from these $n$-point functions
should also contain these information. Using this procedure we
deduce the position dependent counterterms in our problem. We should
mention that most of the authors use the free counterterms, by which
we mean the ones that are relevant to the free cases with no
nontrivial BCs, and are obviously position independent. However, the
dependency of the counterterms on the distance between the plates
has been noted in some references such as
\cite{Albuquerque.,Fosco.}. However, these authors use free
counterterms in the space between the plates and place additional
surface counterterms at the boundaries.
\begin{figure}
    \hspace{-1cm} \includegraphics[width=6.5cm]{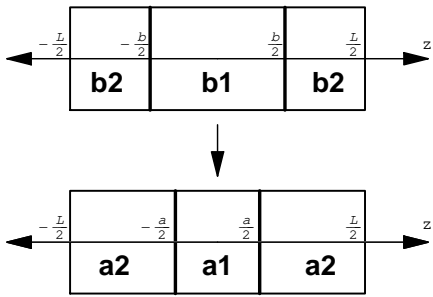} \hspace{0cm}
\caption{\label{box.figs.} The geometry of the two different
configurations whose energies are to be compared. The labels $a1$,
$b1$, etc. denote the appropriate sections in each configuration
separated by plates. The left configuration is denoted by `A'
configuration, and the right one by `B' configuration. }
\end{figure}
\par
Another important part of our calculation is using a method to
remove the divergences without resorting to any analytic
continuation. In fact, we subtract two different configurations with
similar nature. This subtraction scheme is based on the Boyer's
subtraction scheme and it can regularize the infinities and help us
to remove them without using any analytic
continuation\,\cite{boyer.}. This method has been used in many
previous works\,\cite{Lukosz., madad.,phys.rep.353}. We place the
two infinite parallel plates (with distance $a$) within two other
plates (with distance $L > a$). We then construct a similar
configuration of plates with distances $L > b$. We then subtract the
Casimir energies of these two configurations. Finally in order to
obtain the Casimir energy for the original configuration we let $L$
and then $b$ go to infinity. Therefore, the Casimir energy is now
defined by,
\begin{eqnarray}\label{first.def.cas.}
  E_{\mbox{\footnotesize{Cas.}}}=\lim\limits_{b/a \to\infty}\left[\lim\limits_{L/b
  \to\infty}\left(E_{A}-E_{B}\right)\right],
\end{eqnarray}
where
\begin{eqnarray}\label{zeropoint.def.}
  E_{A}=E_{a1}+2E_{a2},\hspace{1.5cm}E_{B}=E_{b1}+2E_{b2},
\end{eqnarray}
and $E_{a1}$, $E_{a2}$, $E_{b1}$ and $E_{b2}$ are the zero point
energies of each region shown in Fig.\,(\ref{box.figs.}).
\par
We have already used this subtraction scheme to calculate the
leading order part of the Casimir energy for a real massive scalar
field and its massless limit with Dirichlet BC for two infinite
parallel plates in arbitrary dimensions in
Refs.\,\cite{reza1.,reza2.}. Therefore, in this paper we only report
its final result for two spatial dimensions. We obtain,
\begin{eqnarray}\label{final.leading.cas.}
 E^{(0)}_{\mbox{\footnotesize{Cas.}}}=-\frac{2L(ma)^{3}}{(4\pi)^{3/2}a^2}
 \sum\limits_{j=1}^{\infty}\frac{K_{3/2}(2amj)}{(a m j)^{3/2}}.
\end{eqnarray}
This expression for the leading order of the Casimir energy of a
massive scalar field with Dirichlet BC in two spatial dimensions is
the same as that reported in Refs.\,\cite{wolf.,phys.rep.353}.
However, contrary to the methods used in
Refs.\,\cite{wolf.,phys.rep.353}, this expression is obtained
without using any analytic continuation techniques. Two important
limits should be considered at this stage. First is the small mass
limit, $m\to 0$, and Eq.\,(\ref{final.leading.cas.}) becomes,
\begin{eqnarray}\label{leading.cas.massless.limit.}
 E^{(0)}_{\mbox{\footnotesize{Cas.}}}=\frac{-L\zeta(3)}{16\pi
 a^{2}},
\end{eqnarray}
where $\zeta(s)$ denotes the zeta function. This expression for the
leading order Casimir energy of a massless scalar field is also the
same as reported in Refs.\,\cite{wolf.,phys.rep.353}. Second is the
large mass limit, $ma\gg1$, and Eq.\,(\ref{final.leading.cas.})
becomes,
\begin{eqnarray}\label{leading.cas.large.mass.limit.}
 E^{(0)}_{\mbox{\footnotesize{Cas.}}}=\frac{-L}{8\pi a^{2}}(am)e^{-2ma}.
\end{eqnarray}
In fact, in this limit the value of the Casimir energy decreases
exponentially with increasing $ma$, and this is again the same as
the previously reported result\,\cite{wolf.,phys.rep.353}.
\par
In section 2, we first calculate the first order radiative
correction to the Casimir energy for this problem. We then plot all
of the result for the massive and massless cases. In section 3, we
summarize and discuss our results.

\section{First-order Radiative Correction}
In this section we first calculate the leading order radiative
correction to the Casimir energy for a massive scalar field within
$\lambda\phi^{4}$ theory with Dirichlet BC in $2+1$ dimensions using
the renormalized perturbation theory. As mentioned in the
Introduction and also Refs.\,\cite{reza1.,reza2.}, the counterterms
are computed from the appropriate $n$-point functions which, in the
presence of the nontrivial BCs, are naturally position dependent.
The renormalization procedure, the deduction of the counterterms,
and the final general form of the first order correction to the
Casimir energy for each region have been completely discussed in
Refs.\,\cite{reza1.,reza2.}. Therefore, in this paper we use only
the conclusions: The general expression for the first order
radiative correction term to the Casimir energy is,
\begin{eqnarray}\label{R.C. def.}
 E^{(1)}_{a1}=\frac{-\lambda}{8}\int_{a1}G^{2}_{a1}(x,x)d^{2}\textbf{x},
\end{eqnarray}
where $G_{a1}(x,x')$ is the propagator of a real scalar field in
region $a1$ in two spatial dimensions. After
the usual wick rotation, the expression for the Green's function or
the propagator $G_{a1}(x,x')$ in three-dimensional Euclidean space
is,
\begin{widetext}
\begin{eqnarray}\label{Green.function.def.}
  G_{a1}(x,x')=\frac{2}a\int\frac{d^{2} k}{(2\pi)^{2}} \sum_{n=1}
  \frac{e^{-\omega(t-t')}e^{-i{\small
  \mathbf{k}^{\bot}}.({\mathbf{x}}^{\bot}-{\mathbf{x'}}^{\bot})}
  \sin\left[k_{a1,n}(z+\frac{a}{2})\right]
\sin\left[k_{a1,n}(z'+\frac{a}{2})\right]}{k^2+k_{a1,n}^2+m^2+i\epsilon},
\end{eqnarray}
where $k_{a1,n}=n\pi/a$, $x=(t,\mathbf{x})$, and
$k=(\omega,\mathbf{k}^{\perp})$. Using Eqs.\,(\ref{R.C.
def.},\ref{Green.function.def.}) and performing the spatial
integration we obtain
\begin{eqnarray}\label{R.C.replacement.}
  E^{(1)}_{a1}&=&\frac{-\lambda}{8}\Bigg[\frac{4}{a^2}\sum_{n,n'=1}\int\frac{d^{2} k}{(2\pi)^{2}}
  \frac{1}{k^2+k_{a1,n}^2+m^2+i\epsilon}\int\frac{d^{2}
  k'}{(2\pi)^{2}}
  \frac{1}{k'^2+k_{a1,n'}^2+m^2+i\epsilon}\bigg[\frac{a
  L}{4}\big(1+\frac{1}{2}\delta_{n,n'}\big)\bigg]\Bigg]\nonumber\\
  &=&\frac{-\lambda L}{32\pi^{2}a}\Bigg[\sum_{n,n'=1}\int_{0}^{\infty}dk
  \frac{k}{k^2+k_{a1,n}^2+m^2+i\epsilon}\int_{0}^{\infty}dk'
  \frac{k'}{k'^2+k_{a1,n'}^2+m^2+i\epsilon}\nonumber\\&&\hspace{1cm}
  +\frac{1}{2}\sum_{n=1}\bigg(\int_{0}^{\infty}dk
  \frac{k}{k^2+k_{a1,n}^2+m^2+i\epsilon}\bigg)^{2}\Bigg].
\end{eqnarray}\end{widetext}
All of the integrals in Eq.\,(\ref{R.C.replacement.}) are
logarithmically divergent, and we make them dimensionless by
multiplying appropriate factors of $a$. Then, we use cutoff
regularization  for each integral, and expand the results in the
limit in which the cutoffs go to infinity as follows,\vspace{0.5cm}
\begin{eqnarray}\label{calculation.of.integration.}
  \int_{0}^{\Lambda}\frac{dk
  k}{k^2+\omega^{2}}=\frac{1}{2}\ln(k^2+\omega^2)
  \Big|_{0}^{\Lambda}\buildrel {\Lambda  \to\infty} \over
 \longrightarrow \ln\Lambda-\ln\omega.
\end{eqnarray}
Using Eq.\,(\ref{calculation.of.integration.}) for each integral in
Eq.\,(\ref{R.C.replacement.}) we obtain,
\begin{widetext}
\begin{eqnarray}\label{R.C.after.integration.}
  E^{(1)}_{a1}={\frac{-\lambda
  L}{32\pi^{2}a}}\Bigg[\sum_{n,n'=1}\Big(\ln\Lambda_{a1}-\ln\omega'_{a1,n}\Big)\Big(\ln\Lambda_{a1}-\ln\omega'_{a1,n'}\Big)
  +\frac{1}{2}\sum_{n=1}\Big(\ln\Lambda_{a1}-\ln\omega'_{a1,n}\Big)^{2}\Bigg],
\end{eqnarray}
where $\omega'^{2}_{a1,n}=(n\pi)^{2}+m^{2}a^{2}$, and $\Lambda_{a1}$
is a cutoff in the upper limit of the integrals in
Eq.\,(\ref{R.C.replacement.}). The terms related to other regions in
Fig.\,(\ref{box.figs.}) are calculated analogously. Now, for the
calculation of the Casimir energy in Eq.\,(\ref{first.def.cas.}), we
have four similar terms which should be subtracted from each other.
By appropriately adjusting each cutoff $\Lambda_{a1}$,
$\Lambda_{a2}$, $\Lambda_{b1}$ and $\Lambda_{b2}$, all of the
infinities cancel due to our box subtraction scheme. We obtain,
\begin{eqnarray}\label{R.C.subtraction.}
  E^{(1)}_{A}-E^{(1)}_{B}&=&E^{(1)}_{a1}+2E^{(1)}_{a2}-E^{(1)}_{b1}-2E^{(1)}_{b2}\nonumber\\ &=&\frac{-\lambda
  L}{32\pi^{2}}\Bigg[\sum_{n,n'=1}\Big(\frac{\ln\omega'_{a1,n}\ln\omega'_{a1,n'}}{a}
  +4\frac{\ln\omega'_{a2,n}\ln\omega'_{a2,n'}}{L-a}
  -\frac{\ln\omega'_{b1,n}\ln\omega'_{b1,n'}}{b}-4\frac{\ln\omega'_{b2,n}\ln\omega'_{b2,n'}}{L-b}\Big)
  \nonumber\\&&\hspace{1cm}+\frac{1}{2}\sum_{n=1}\Big(\frac{\ln^{2}\omega'_{a1,n}}{a}
  +4\frac{\ln^{2}\omega'_{a2,n}}{L-a}-\frac{\ln^{2}\omega'_{b1,n}}{b}-4\frac{\ln^{2}\omega'_{b2,n}}{L-b}\Big)\Bigg].
\end{eqnarray}
Now we can use the Abel-Plana Summation Formula (APSF) which
basically reduces the summations into integrations as follows,
\begin{eqnarray}\label{APSF}
\sum_{n=1}^{\infty}F(n)=\frac{-1}{2}F\left( 0 \right) +
\int_0^\infty dt F\left( t \right)+ i\int_0^\infty dt \frac{F\left(
it \right) - F\left( - it \right)}{e^{2\pi t}- 1}.
\end{eqnarray}
If the summation starts from $n=0$, the sign of the first term
becomes positive. Now by applying the APSF to all of the summations
in Eq.\,(\ref{R.C.subtraction.}) we obtain,
\begin{eqnarray}\label{R.C.abel.plana.employed}
  E^{(1)}_{A}-E^{(1)}_{B}=\frac{-\lambda
  L}{128\pi^{2}}\Bigg[\mathcal{R}(a)+2\mathcal{R}(\frac{L-a}{2})-\{a\to
  b\}\Bigg],\hspace{0.5cm}
\end{eqnarray}
where
\begin{eqnarray}\label{f.definition}
  \mathcal{R}(x)=\frac{1}{x}\Bigg(\frac{-1}{2}\ln
  m^2x^2+\int_{0}^{\infty}\hspace{-0.2cm}dn\ln\big(n^2\pi^2+m^2x^2\big)+B_1(x)\Bigg)^2
  \hspace{-0.3cm}+\frac{1}{2x}\Bigg(\frac{-1}{2}\ln^{2}m^2x^2
  +\int_{0}^{\infty}\hspace{-0.2cm}dn\ln^2\big(n^2\pi^2+m^2x^2\big)+B_2(x)\Bigg),\nonumber\hspace{-1cm}\\
\end{eqnarray}\end{widetext}
and $B_1(x)$ and $B_{2}(x)$ are the branch-cut terms of the
Abel-Plana summation formula and are calculated in
Appendix\,\ref{Appendix.1}. Both of these two types of branch-cut
terms are finite for $m\neq0$. However, other integral terms which
appear in Eq.\,(\ref{R.C.abel.plana.employed}) are divergent. At
this stage our main purpose is to regularize these terms and show
how they cancel each other, again due to our box subtraction scheme.
\par
To regularize the integrals in Eq.\,(\ref{R.C.abel.plana.employed}),
we set separate cutoffs, denoted again by $\Lambda$s, for the upper
limits of each integral. After the integrations, we expand the
results in the limit $\Lambda\to\infty$. Now, by appropriate
adjustment of the $\Lambda$s, all of the divergent terms which
depend on the cutoffs $\Lambda$s, cancel in
Eq.\,(\ref{R.C.abel.plana.employed}), due to our box subtraction
scheme. Below, we present the details of these cancelations for both
types of integrals.  For the first type we have,\vspace{-2cm}
\begin{widetext}
\begin{eqnarray}\label{first.integral.result.}
  \int_{0}^{\Lambda}dn\ln\big(n^2\pi^2+m^2a^2\big)&=&\frac{ma}{\pi}\int_{0}^{\Lambda}dN\bigg(\ln\big(N^2+1\big)+\ln\big(m^2
  a^2\big)\bigg)\nonumber \\
  &=&\frac{ma\Lambda}{\pi}\Big[-2+\ln(m^2a^2)+\ln(1+\Lambda^2)+\frac{2}{\Lambda}\arctan\Lambda\Big]
  \nonumber\\ &&\buildrel {\Lambda  \to\infty} \over
  \longrightarrow \frac{ma}{\pi}\bigg(-2+\ln \big(m^2a^2\Lambda^2\big)\bigg)\Lambda+ma-\frac{ma}{\pi\Lambda}
  +\mathcal{O}\Big(\frac{1}{\Lambda}\Big)^3 \longrightarrow ma,
\end{eqnarray}
where in the first line we have used the following change of
variable $N=n\pi/ma$. Therefore, only the finite terms
$\{am,(L-a)m,bm,(L-b)m\}$ remain for the first type of integrals.
For the second type of integrals we have,
\begin{eqnarray}\label{second.integral.result.}
  \int_{0}^{\Lambda}dn\ln^2\big(n^2\pi^2+m^2a^2\big)
  &=&\frac{ma}{\pi}\int_{0}^{\Lambda}dN\Big[\ln(a^2m^2)+\ln(N^2+1)\Big]^2
  \nonumber\\ &=&\frac{ma}{\pi}\int_{0}^{\Lambda}dN\Big[\ln^2(a^2m^2)+2\ln(a^2m^2)\ln(N^2+1)+\ln^2(N^2+1)\Big]
  \nonumber\\
  &=&\frac{ma}{\pi}\ln^2(a^2m^2)\Lambda
  +\frac{2ma}{\pi}\ln(a^2m^2)
  \underbrace{\big\{-2N+2\arctan(N)+N\ln(N^2+1)\big\}\bigg|_0^{\Lambda}}_{\mathcal{K}(\Lambda)}
  \nonumber\\&&+\frac{ma}{\pi}\int_{0}^{\Lambda}dN\ln^2(N^2+1),
\end{eqnarray}
\end{widetext}
where $\mathcal{K}(\Lambda)$ in the limit $\Lambda\to\infty$ is,
\begin{eqnarray}\label{k.lambda.}
  \hspace{-0.5cm}\mathcal{K}(\Lambda)\buildrel  { \Lambda\to\infty}\over \longrightarrow
  (-2+\ln\Lambda^2)\Lambda+\pi-\frac{1}{\Lambda}+\mathcal{O}(\frac{1}{\Lambda})^2\longrightarrow\pi.
\end{eqnarray}
The first term in the right hand side of
Eq.\,(\ref{second.integral.result.}) in line three is divergent. The
second integration is similar to the first type of integral terms
which were considered above. The divergent terms in the sum of these
two terms are removed by choosing appropriate adjustment of
$\Lambda$s and using the subtraction scheme indicated in
Eq.\,(\ref{R.C.abel.plana.employed}), and only finite terms remain.
The third term in the right hand side of
Eq.\,(\ref{second.integral.result.}) and its counterparts in the
other regions are also divergent and their calculations are very
difficult. However, if we let first the cutoffs go to infinity, one
can show that they exactly cancel each other in the box subtraction
scheme. Therefore, the only contributions coming from this term is,
\begin{eqnarray}\label{second.integral.final.result.}
  \int_{0}^{\infty}dn\ln^2\big(n^2\pi^2+m^2a^2\big)\longrightarrow4ma\ln ma.
\end{eqnarray}
Using
Eqs.\,(\ref{R.C.abel.plana.employed},\ref{first.integral.result.},\ref{second.integral.final.result.}),
we have
\begin{widetext}
\begin{eqnarray}\label{R.C.integrals.caculated.}
E^{(1)}_{A}-E^{(1)}_{B}&=&\frac{-\lambda
  L}{128\pi^{2}}\Bigg[\frac{1}{a}\Big(a^2m^2+B_1^2(a)-ma\ln(m^2a^2)+2ma B_1(a)-B_1(a)\ln(m^2a^2)
  +\frac{1}{2}\big(4ma\ln
  ma\big)+\frac{1}{2}B_2(a)\Big)\nonumber\\ &&+
  \frac{2}{L-a}\Big(\frac{(L-a)^2m^2}{4}+B_1^2(\frac{L-a}{2})
  -\frac{m(L-a)}{2}\ln(\frac{m^2(L-a)^2}{4})
  +m(L-a)B_1(\frac{L-a}{2})\nonumber\\ &&-B_1(\frac{L-a}{2})\ln\big(\frac{(L-a)^2m^2}{4}\big)
  +m(L-a)\ln\big(\frac{m(L-a)}{2}\big)
  +\frac{1}{2}B_2(\frac{L-a}{2})\Big)\nonumber\\ &&-\big\{a\to b\big\}\Bigg].
\end{eqnarray}
\end{widetext}
There many internal cancelations in the above expressions. After
these cancelations only the  branch-cut terms remain. By using the
values of the branch-cut terms obtained in the
Appendix\,\ref{Appendix.1}, we can write an explicit expression for
the lowest order radiative correction to the Casimir energy in terms
of parameters $m,a,\frac{L-a}{2},b$ and $\frac{L-b}{2}$. As stated
in the Eq.\,(\ref{first.def.cas.}), first the limit $L/b\to\infty$
should be calculated and then $b/a\to\infty$. In these limits all of
the terms which depend on $L$ and $b$ disappear from our expression
and only the terms which depend on the distance of the original
plates\,($a$) remain. Our final result is,
\begin{eqnarray}\label{R.C.final.result}
E^{(1)}_{\mbox{\footnotesize{Cas.}}}&=&\frac{-\lambda
  L}{128\pi^{2}a}\nonumber\\ &&\hspace{-1cm}\times\Bigg[\Big(am+\ln\big(1-e^{-2am}\big)\Big)^2-m^2a^2
  -\gamma\ln\big(1-e^{-2am}\big)
  \nonumber\\ &&\hspace{-0.5cm}+\sum\limits_{j=1}^{\infty}\frac{e^{-2amj}}{j}
  \Big(\ln\big(maj\big)-e^{4maj}\Gamma(0,4maj)\Big)\Bigg],
\end{eqnarray}
\begin{figure}
    \hspace{-1cm} \includegraphics[width=7.5cm]{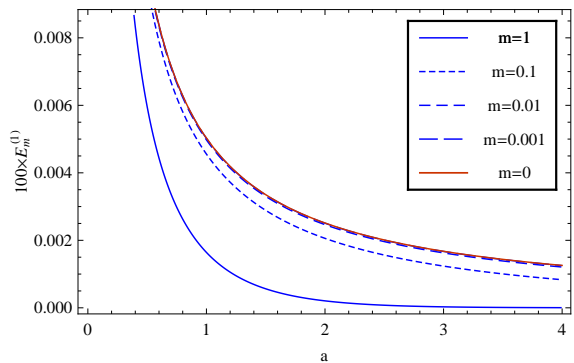}
    \caption{\label{E.Cas.massles.case.final}
    The first order radiative corrections to the Casimir energy for massive and massless scalar
    fields in two spatial dimensions are plotted as a function of the distance between
    the lines\,($a$), within the $\lambda \phi^4$ theory for $\lambda=0.1$.
    The numerical values for the plots have been multiplied by a factor of $100$,
    in order to make their absolute values comparable to the zero order terms shown
    in Fig.\,(\ref{E.Cas.all.}). In this figure we have shown the sequence of
    plots for $m=\{1,0.1,0.01,0.001,0\}$. It is apparent that the
    sequence of the massive cases converges rapidly to the massless case and there is an insignificant difference between
    the figures of the massive cases for $m<0.01$, and the massless case.     }
\end{figure}
where $\gamma$ is the Euler-Mascheroni constant, and
$\Gamma(\alpha,x)$ is the incomplete Gamma function. Our result
differs from the previously reported result\,\cite{cavalcanti.},
since they use the free counterterms, and we have used the ones
dictated by the Green's function appropriate for this problem.
\par
To calculate the massless limit, we go back to the original
expression given in Eq.\,(\ref{R.C.integrals.caculated.}). The
direct calculation of the massless case is extremely difficult. We
use $m$ as a regulator for this limit. However, multitude of
difficulties appear. These difficulties are partly due to the fact
that the branch-cut terms are also divergent in the limit $m\to 0$.
Fortunately, there is no essential singularity and we obtain
\begin{eqnarray}\label{massless.limit1.}
&&\hspace{-1.1cm}E^{(1)}_{\mbox{\footnotesize{Cas.}}}=\frac{-\lambda
  L}{128\pi^{2}a}\Bigg[\Big(am+B_{1}(a)\Big)^2-m^2a^2\nonumber\\&&
  \hspace{1.6cm}-2B_{1}(a)\ln(ma)+\frac{1}{2}B_{2}(x)\Bigg]
  \nonumber\\ &\buildrel {m  \to 0} \over
  \longrightarrow&\frac{-\lambda
  L}{128\pi^{2}a}\Bigg[\ln^{2}(2ma)-2ma\int_{1}^{\infty}dN\frac{\ln(N^{2}-1)}{e^{2maN}-1}\Bigg],\nonumber\\&&
\end{eqnarray}
where in the second line we have used the small mass limit of
$B_{1}$,
\begin{eqnarray}\label{B1.massless.limit.}
\hspace{-0.3cm}B_{1}(a)=\ln(1-e^{-2ma})\buildrel {m \to 0} \over
  \longrightarrow\ln(2ma)-ma+\mathcal{O}(m^2),
\end{eqnarray}
and used a suitable change of variables for $B_{2}$ which leads to
some cancelations, and we have ignored terms of $\mathcal{O}(m^2)$.
In the above expression all of the infinities cancel and we finally
obtain the following finite result,
\begin{eqnarray}\label{massless.limit.Cas.}
E^{(1)}_{\mbox{\footnotesize{Cas.}}}\buildrel {m \to 0} \over
  \longrightarrow\frac{-\lambda
  L}{128\pi^{2}a}\big(-0.6349208\big).
\end{eqnarray}
As shown in Fig.\,(\ref{E.Cas.massles.case.final}), the sequence of
plots of the massive cases converges rapidly to the massless limit.
It is obvious that the massless limit is finite, exactly as we have
obtained, and as expected on physical grounds.
\begin{figure}
    \hspace{-1cm} \includegraphics[width=8cm]{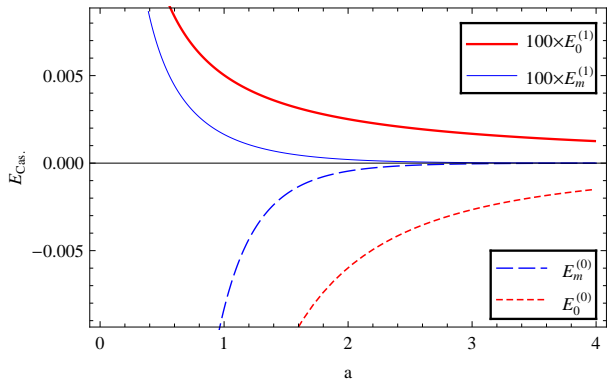}
    \caption{\label{E.Cas.all.}   The leading
    term for the Casimir energy and its first order radiative
    correction (multiplied by a factor of $100$) in two spatial
    dimensions, are plotted as a function of the distance between the
    lines\,($a$) for a massive ($m=1$) and a massless scalar fields
    for $\lambda=0.1$\,. The correction terms are always positive.  }
\end{figure}
\par
In Fig.\,(\ref{E.Cas.all.}) all of the values for the zero order and
the first order radiative correction to the Casimir energy for a
massive\,($m=1$) and massless scalar fields are plotted. We should
mention that the correction terms are positive and their values are
approximately $100$ times smaller than their zero-order
counterparts.

\section{Conclusion}
In this paper, the first order radiative correction to the Casimir
energy with Dirichlet BC for two infinite parallel plates in two
spatial dimensions has been calculated by a systematic approach to
the renormalization program. This program automatically yields
position dependent counterterms. Moreover, we used the Boyer's
subtraction scheme which eliminates the need to use any analytic
continuation techniques. The final results for the radiative
correction of the Casimir energy for a massive and massless scalar
fields are different from the reported results in the previous
papers\,\cite{cavalcanti.}. We believe that this difference is due
to the use of different renormalization programs. It is important to
note that our result for the massless case is finite, in sharp
contrast to the previously reported result\,\cite{cavalcanti.}.

\appendix
\section{Calculation of the Branch-cut terms} \label{Appendix.1}
In this Appendix we present the calculation of two types of branch-cut terms which appear in
the calculation of the first order radiative correction to the Casimir
energy. We start with the first type of the
branch-cut term which is denoted by $B_{1}$ in the main text. We have for $B_{1}(a)$,
\begin{widetext}
\begin{eqnarray}\label{calculation.branch.1}
  B_{1}(a)&=&i\int_{\frac{ma}{\pi}}^{\infty}dn
  \frac{\ln\big((in)^2\pi^2+m^2a^2\big)-\ln\big((-in)^2\pi^2+m^2a^2\big)}{e^{2\pi
  n}-1}\nonumber\\&=& i\int_{\frac{ma}{\pi}}^{\infty}dn\frac{\ln\big(
  e^{i\pi}n^2\pi^2+m^2a^2\big)-\ln\big(e^{-i\pi}n^2\pi^2+m^2a^2\big)}{e^{2\pi
  n}-1}= -2\pi\int_{\frac{ma}{\pi}}^{\infty}\frac{dn}{e^{2\pi
  n}-1}=\ln\big(1-e^{-2am}\big).
\end{eqnarray}
The values of other branch-cut terms can be easily written only by
the appropriate replacement in the argument of $B_{1}$. Analogous
process is repeated for calculation of the second type of the
branch-cut term. So, for $B_{2}(a)$ we have,
\begin{eqnarray}\label{calculation.branch.2}
  B_{2}(a)&=&i\int_{\frac{ma}{\pi}}^{\infty}dn\frac{\ln^2\big((in)^2\pi^2+m^2a^2\big)
  -\ln^2\big((-in)^2\pi^2+m^2a^2\big)}{e^{2\pi n}-1}\\&=&
  i\int_{\frac{ma}{\pi}}^{\infty}dn\frac{\bigg(i\pi+\ln\big(n^2\pi^2-m^2a^2\big)\bigg)^2
  -\bigg(-i\pi+\ln\big(n^2\pi^2-m^2a^2\big)\bigg)^2}{ e^{2\pi n}-1}
  =-4\pi\int_{\frac{ma}{\pi}}^{\infty}dn\frac{\ln\big(n^2\pi^2-m^2a^2\big)}{e^{2\pi
  n}-1}\nonumber.
\end{eqnarray}\end{widetext}
This integral can not be performed in closed form. Expanding the
denominator of the integrand we obtain,
\begin{eqnarray}\label{ calculation.branch.2-edameh.}
  &\hspace{-1cm}\dd B_{2}(a)=-4\pi\sum\limits_{j=1}^{\infty}\int_{\frac{ma}{\pi}}^{\infty}dn
  e^{-2\pi n j}\ln\big(n^2\pi^2-m^2a^2\big)\nonumber\\  &\hspace{-0.3cm}\dd=
  -4\pi\Bigg\{\frac{\gamma}{2\pi}\ln\big(1-e^{-2am}\big)
  +\sum\limits_{j=1}^{\infty}\frac{e^{-2amj}}{2\pi
  j}\nonumber\\  &\hspace{2.3cm}\dd\times\Bigg(\ln\big(am/j\big)+e^{4amj}\Gamma\big(0,4amj\big)\Bigg)\Bigg\},\nonumber\\
\end{eqnarray}
where $\Gamma(\alpha,x)$ is the incomplete gamma function and in our
case we have,
\begin{eqnarray}\label{gamma.besselk.}
  \Gamma\big(0,x\big)=-e^{-x/2}\sqrt{\frac{x}{\pi}}\partial_{\nu}K_{\nu}\big(x/2\big)\Big|_{\nu=-1/2}.
\end{eqnarray}
Both the first and second type of the branch-cut terms are finite
for $m\neq0$ and also their values go to zero when their arguments
tend to infinity. So, when the limit $L/a\to \infty$ and
$b/a\to\infty$ are taken, the contributions of these two branch-cut
terms go to zero and therefore, only the branch-cut terms which
depend on the original distance $a$ remain.

\setcounter{equation}{0}
\renewcommand{\theequation}{\Alph{section}.\arabic{equation}}

\section*{Acknowledgement} We would like to thank the Research Office
of the Shahid Beheshti University for financial support.



\begin{thebibliography}{9}
\bibitem{plunien.}
     G. Plunien, B. Muller, and W. Greiner, \emph{The Casimir effect, Phys. Rep.} \textbf{134}, 87 (1986).
\bibitem{mostepanenko.}
     V. M. Mostepanenko and N. N. Trunov, \emph{The Casimir effect and its applications}, Clarendon, Oxford,
     (1997).
\bibitem{phys.rep.353}
     M. Bordag, U. Mohideen, and V. M. Mostepanenko, \emph{New
     developments in the Casimir effect, Phys. Rep.} \textbf{353}, 1 (2001). [{\tt arXiv:quant-ph/0106045}]
\bibitem{h.b.g.}
     H.B.G.~Casimir and D.~Polder, \emph{The Influence of Retardation on the London-van der Waals
     Forces, Phys. Rev.} \textbf{73} 360 (1948);\\
     H. B. G. Casimir, \emph{On the attraction between two perfectly conducting plates, Proc. Kon. Aa. Wet.} \textbf{51}, 793 (1948).
\bibitem{milton.paper.}
     K. A. Milton,  \emph{The Casimir Effect: Physical Manifestations of Zero Point Energy},
     Invited Lectures at 17th Symposium on Theoretical Physics, Seoul National University, Korea, June 29-July 1,
     (1998). [{\tt arXiv:hep-th/9901011}]
\bibitem{milton.book.}
     K. A. Milton, \emph{The Casimir Effect: Physical Manifestations of Zero-Point Energy}, (World
     Scientific Publishing Co. Pte. Ltd. 2001).
\bibitem{sparnaay}
     M. J. Sparnaay, \emph{Measurements of attractive forces between flat plates, Physica} \textbf{24}, 751 (1958).
\bibitem{Bressi.}
     G. Bressi, G. Carugno, R. Onfrio, G. Ruoso, \emph{Measurement of the Casimir Force between Parallel
     Metallic Surfaces, Phys. Rev. Lett.} \textbf{88}, 041804 (2002).
\bibitem{Lamoreaux.decca1.decca2.}
     S. K. Lamoreaux, \emph{Demonstration of the Casimir Force in the $0.6$ to 6 $\mu m$ Range,
     Phys. Rev. Lett.} \textbf{78}, 5 (1997);\\
     R.S. Decca, D. L\'{o}pez, H.B. Chan, E. Fischbach, D.E. Krause, and C.R. Jamell, \emph{Constraining New
     Forces in the Casimir Regime Using the Isoelectronic Technique, Phys. Rev. Lett.} \textbf{94}, 240401 (2005);\\
     R.S. Decca, D. L\'{o}pez, E. Fischbach, G.L. Klimchitskaya, D.E. Krause, and V.M. Mostepanenko,
     \emph{Precise comparison of theory and new experiment for the Casimir force leads to stronger constraints on
     thermal quantum effects and long-range interactions, Annals of Physics}, \textbf{318}, 37 (2005).
\bibitem{wolf.}
     J. Ambj{\o}rn and S. Wolfram, \emph{Properties of the vacuum, 1. Mechanical and thermodynamic,
     Ann. Phys. (N.Y.)} \textbf{147}, 1 (1983).
\bibitem{svaiters}
     N. F. Svaiter and B. F. Svaiter, \emph{Casimir effect in a d-dimensional flat space-time and the cutoff method, J. math.
     Phys.} \textbf{32}, 175 (1991);\\
\bibitem{cognola.}
     G. Cognola, E. Elizalde, and K. Kiresten, \emph{Casimir energies for
     spherically symmetric cavities, J. Phys. A: Math. Gen.}
     \textbf{34}, 7311 (2001).
\bibitem{mode.sum.}
     A. Romeo, K. A. Milton, \emph{Casimir Energy for a Purely Dielectric Cylinder by the Mode Summation Method, Phys.Lett. B}\textbf{621}, 309
     (2005);\\
     K. A. Milton, A. V. Nesterenko, V. V. Nesterenko, \emph{Mode-by-mode summation for the zero point
     electromagnetic energy of an infinite cylinder, Phys.Rev.D}\textbf{59}, 105009
     (1999);\\
     I. H. Brevik, V. V. Nesterenko, I. G. Pirozhenko, \emph{Direct mode summation
     for the Casimir energy of a solid ball, J. Phys. A}\textbf{31}, 8661
     (1998);
\bibitem{nesterenko.piro.}
     V. V. Nesterenko, I. G. Pirozhenko, \emph{Spectral Zeta Functions for
     a Cylinder and a Circle, J. Math. Phys.} \textbf{41} 4521
     (2000).
\bibitem{green.func.}
     K. A. Milton, L. L. Deraad, and J. Schwinger, \emph{Casimir self-stress on a perfectly conducting spherical shell, Ann. Phys. (N.Y.)} \textbf{115}, 388 (1978).
\bibitem{multiple.scatt.exp.}
     R. Balian, and B. Duplantier, \emph{Electromagnetic waves near perfect conductors. II. Casimir effect, Ann. Phys. (N.Y.)} \textbf{112},
     165 (1978).
\bibitem{heat.kernel.}
     T. P. Branson and P. B. Gilkey, \emph{The Asymptotics of The Laplacian on a Manifold with Boundary, Commun. Partial Differential
     Eqs.}\textbf{15}, 245 (1990);\\
     M. Bordag and K. Kiresten, \emph{Heat kernel Coefficients and Divergencies of the Casimir Energy for the Dispersive Sphere, Int. J. Mod. Phys. A}\textbf{17}, 813
     (2002).
\bibitem{onetwo.loop.}
     M. Bordag, D. Robaschik and E. Wieczorek, \emph{Quantum field Theoric treatment of the Casimir effect, Ann.
     Phys.} (NY) \textbf{165}, 192 (1985);\\
     M. Bordag, and J. Lindig, \emph{Radiative correction to the Casimir
     force on a sphere, Phys. Rev. D} \textbf{58}, 045003
     (1998); [{\tt arXive:hep-th/9801129}]\\
     D. Robaschik, K. Scharnhorst and E. Wieczorek, \emph{Radiative
     corrections to the Casimir pressure under he influence of
     temperature and external fields, Ann. Phys. (NY)} \textbf{174},
     401 (1987);\\
     M. Bordagand K. Scharnhorst, \emph{$O(\alpha)$ Radiative Correction to the Casimir Energy for
     Penetrable Mirrors, Phys. Rev. Lett.} \textbf{81}, 3815 (1998); [{\tt arXive:hep-th/9807121}]\\
     S. S. Xue., \emph{Casimir effect of scalar field on S(n-1) manifold, Commun. Theor. phys. (Wuhan)} \textbf{11}, 243 (1989).
\bibitem{6.reza.3}
     F. Ravndal and J.B. Thomassen, \emph{Radiative corrections to the Casimir energy and effective
     field theory, Phys. Rev. D} \textbf{63}, 113007 (2001).
\bibitem{7va8.reza.3}
     X. Kong and F. Ravndal, \emph{Radiative Corrections to the Casimir Energy,
     Phys. Rev. Lett.} \textbf{79}, 545 (1997);\\
     K. Melnikov, \emph{Radiative corrections
     to the Casimir force and effective field theories, Phys. Rev. D} \textbf{64}, 045002 (2001).
\bibitem{cavalcanti.}
     R.M. Cavalcanti, C. Farina, and  F.A. Barone, \emph{Radiative
     Corrections to Casimir Effect in the $\lambda \phi^4$ Model},
     [{\tt arXive:hep-th/0604200}] (2006);\\
     F. A. Barone, R. M. Cavalcanti, and C. Farina, \emph{Radiative
     corrections to the Casimir effect for the massive scalar field},
     [{\tt arXive: hep-th/0301238v1}](2003);\\
     F. A. Barone, R. M. Cavalcanti, and C. Farina, \emph{Radiative
     corrections to the Casimir effect for the massive scalar field,
     Nucl. Phys. Proc. Suppl.} \textbf{127}, 118 (2004). [{\tt arXive:
     hep-th/0306011v2}]
\bibitem{graham.IJMP.}
     N. Graham, R. Jaffe, H. Weigel, \emph{Casimir Effects in Renormalizable Quantum Field Theories, Int. J. Mod. Phys. A} \textbf{17}, 846 (2002).
\bibitem{graham.jaffe.}
     N. Graham, R.L. Jaffe, V. Khemani, M. Quandt, M. Scandurra, H. Weigel, \emph{Calculating vacuum energies in renormalizable
     quantum field theories: A new approach to the Casimir problem, Nucl.
     Phys. B}\textbf{645}, 49 (2002);\\
     N. Graham, R.L. Jaffe, V. Khemani, M. Quandt, O. Schr\"{o}der, H. Weigel, \emph{The Dirichlet Casimir problem, Nucl. Phys. B}\textbf{677}, 379 (2004).
\bibitem{milton.potential.}
     K. A. Milton, \emph{The Casimir effect: recent controversies and
     progress, J. Phys. A: Math. Gen.} \textbf{37}, 209 (2004).
\bibitem{vega...}
     H.J. Vega, \emph{Two-Loop Quantum Correction to the Soliton Mass in
     Two-Dimensional Scalar Field Theories, Nucl. Phys. B}\textbf{115}, 411 (1976);\\
     M.A. Lohe, D.M. O'Brien, \emph{Soliton mass corrections and explicit
     models in two dimensions, Phys. Rev. D} \textbf{23} 1771 (1981);\\
     N. Graham, R.L. Jaffe,\emph{Fermionic one-loop corrections to
     soliton energies in 1+ 1 dimensions, Nucl. Phys. B}\textbf{549} 516
     (1999);\\
     A. A. Izquierdo, W.G. Fuertes, M.A. Gonz\'{a}lez Le\'{o}n, J.M.
     Guilarte, \emph{Generalized zeta functions and one-loop corrections
     to quantum kink masses, Nucl. Phys. B}\textbf{635} 525 (2002); \\
     A. Rebhan, P. van Nieuwenhuizen, R. Wimmer, \emph{The anomaly in the
     central charge of the supersymmetric kink from dimensional
     regularization and reduction, Nucl. Phys. B}\textbf{648}, 174 (2003); \\
     A.A. Izquierdo, W.G. Fuertes, M.A. Gonz\'{a}lez Le\'{o}n, J.M.
     Guilarte, \emph{One-loop corrections to classical masses of kink
     families, Nucl. Phys. B}\textbf{681}, 163 (2004).
\bibitem{dashen.}
     R.F. Dashen, B. Hasslacher, A. Neveu, \emph{Nonperturbative methods
     and extended-hadron models in field theory. II. Two-dimensional
     models and extended hadrons, Phys. Rev. D}\textbf{10}, 4130 (1974).
\bibitem{yamagishi.}
     H. Yamagishi, \emph{Soliton Mass Distributions in ($1+1$)-Dimensional Supersymmetric Theories, Phys. Lett.
     B}\textbf{147}, 425 (1984).
\bibitem{bender.milton.}
     C. M. Bender and K. A. Milton, \emph{Scalar Casimir effect for a
     D-dimensional sphere, Phys. rev. D}\textbf{50}, 6547 (1994).
\bibitem{reza1.}
     R. Moazzemi, M. Namdar, and S. S. Gousheh,\emph{The Dirichlet Casimir effect for $\phi^4$ theory in (3 + 1)
     dimensions: a new renormalization approach, JHEP} \textbf{09}, 029
     (2007), [{\tt arXiv:hep-th/0708.4127v1}].
\bibitem{reza2.}
     R. Moazzemi, S. S. Gousheh, \emph{A
     new renormalization approach to the Dirichlet Casimir effect for $\phi^4$ theory in 1+1 dimensions, Phys. Lett. B}
     \textbf{658}, 255 (2008), [{\tt arXiv:hep-th/0708.3428v2}].

\bibitem{peskin.book.}
     M. E. Peskin and D. V. Schroeder, \emph{An Introduction to Quantum Field Theory},
     Addison-Wesley Pub. Co. (1995).

\bibitem{Albuquerque.}
     L. C. de Albuquerque, \emph{Casimir pressure at two loops and soft boundaries at finite temperature, Phys. Rev. D} \textbf{55}, 7754
     (1997).
\bibitem{Fosco.}
     C.D. Fosco, N.F. Svaiter, \emph{Finite size effects in the
     anisotropic $\lambda(\phi_{1}^{4}+\phi_{2}^{4})/4!$ model, J. Math.
     Phys.} \textbf{42}, 5185 (2001).
\bibitem{boyer.}
     T. H. Boyer, \emph{Quantum Electromagnetic Zero-Point Energy of a
     Conducting Spherical Shell and the Casimir Model for a Charged Particle, Phys. Rev.} \textbf{174}, 1764 (1968).
\bibitem{Lukosz.}
     W. Lukosz, \emph{Electromagnetic zero-point energy and radiation pressure for a rectangular cavity, Physica}, \textbf{56}, 109 (1971).
\bibitem{madad.}
     M. A. Valuyan, R. Moazzemi, and S. S. Gousheh, \emph{A direct approach to the electromagnetic Casimir energy in a rectangular waveguide, J. Phys. B: At. Mol. Opt.
     Phys.} \textbf{41}, 145502 (2008).




\end{thebibliography}
 \end{document}